\documentstyle[12pt]{article}

\begin{document}

\rightline{CfPA-TH-93-01}
\rightline{YCTP-P44-92}
\rightline{Feb.~1993}
\rightline{Revised Mar.~1993}
\rightline{Final revision Aug.~1993}

\vskip .2in

\begin{center}
{\large{\bf Cosmic Variance in CMB Anisotropies: From $1^{\circ}$ to COBE}}
\end{center}

\vskip .1in

\begin{center}
Martin White$^{\dagger}$,
Lawrence M. Krauss$^{\ddagger},$\footnote{Research supported in part by
the NSF,DOE, and TNRLC.  Current address Department of Physics, Case
Western Reserve University, Cleveland, OH 44106-7215}
and Joseph Silk$^{\dagger}$

$^{\dagger}$ {\it Center for Particle Astrophysics and Departments of
Astronomy and Physics, University of California, Berkeley, CA 94720}
$^{\ddagger}$ {\it Center for Theoretical Physics and Dept. of Astronomy,
Sloane Laboratory, Yale University, New Haven, CT 06511}

\end{center}

\vskip .1in

\centerline{ {\bf Abstract} }

\noindent Cosmic Microwave Background (CMB) anisotropies that result from
quantum fluctuations during inflation are explored and the impact of
their ``cosmic variance'' on the ability to use existing data to probe
inflationary models is studied.
We calculate the rms temperature fluctuation, and its cosmic variance, for
a number of experiments and for models with primordial power spectra which
range from $n={1\over 2}$ to $1$.  We find:
(1) cosmic variance obscures the information which can be extracted,
so a comparison of the rms temperature fluctuation on small scales with the
COBE result can fix $n$ to only $\approx\pm 0.2$ at best;
(2) measurements of the rms fluctuation on  $1^{\circ}$ scales may not
allow one to unambiguously infer the tensor contribution to the COBE
anisotropy;
(3) comparison of this contribution with the predictions of inflation are
ambiguous if the quadrupole anisotropy alone is utilized.
We discuss means for minimizing the uncertainty due to cosmic variance in
comparisons between experiments.

\vskip .1in

\noindent {\em Subject headings}: cosmology: cosmic background radiation

\newpage

\noindent{\bf 1. Introduction}

One of the central purposes of measuring CMB anisotropies is to gain
information on the power spectrum of primordial density fluctuations, and
from this, information on the physical processes in the early universe
which may have produced such a spectrum.  The well studied mechanism of
inflation allows an {\it a priori} calculation of the spectrum of fluctuations.
Both scalar and tensor fluctuations arise from calculable quantum fluctuations
in elementary fields during an inflationary phase in the early universe, and
any previously existing density fluctuations are inflated away.
We explore here what constraints on the primordial power spectrum can be
obtained, both in principle and in practice, from the COBE DMR observations in
combination with observations of CMB anisotropies on scales as small as
$1^{\circ}$. Comparison with large scale structure analyses (e.g.
$\delta T/T|_{SW}$ as inferred from POTENT (Bertschinger \& Dekel~1989;
Bertschinger, Gorski \& Dekel~1990) ), could provide useful additional
constraints although at present there are large theoretical and observational
uncertainties.

When comparing theoretical predictions with observations, particular
attention should be paid to the fact that CMB anisotropies resulting from
inflation are stochastic. Because of the quantum nature of the process by which
fluctuations are generated, only observations over an {\it ensemble} of
universes can allow the unambiguous measurement of fundamental
microphysical parameters, even in principle\footnote{We shall discuss
this issue in more  detail later in the text.}. Constrained to observe only
one universe, there remains an irremovable uncertainty in our ability to
relate certain CMB measurements, no matter how precise, to predictions
of inflationary models (such as the normalization and slope of the power
spectrum). For example, inflation predicts a probability distribution for
ratios of the moments of CMB temperature fluctuations (or functions
thereof), and {\em not} the ratios themselves.  This has important
consequences for the normalization of the power spectrum. We term the induced
uncertainty the ``cosmic variance", not to be confused with other
usages of this term. Among our results will be a description of the effect
of this variance on the constraints which can be derived from present and
future observations of the CMB.

\vskip .1in

\noindent{{\bf 2. The Primordial Power Spectrum and Inflation}}

The remarkable observations of primordial anisotropies in the CMB by the
DMR instrument aboard COBE (Smoot et al.~1992) offered cosmologists the
first ``unprocessed" glimpse of what may be the initial conditions for the
present observed Friedman-Robertson-Walker expansion of the universe.
Inferring the primordial power spectrum of fluctuations based on observations
of large scale structure had previously required detailed assumptions about
both the equation of state of matter inside the horizon and the relationship
between the observed structure and the underlying distribution of mass in the
universe. Now model predictions can be normalized at COBE scales,
which represent modes which had not crossed the horizon at recombination and
hence could not have been processed by causal microphysical factors. This
represents a significant theoretical advance. Moreover, the
scales observed by COBE in combination with small scale observations
provide a large lever arm for exploring the scale dependence of the power
spectrum.

A central quantity of interest has been the slope of the power spectrum of
fluctuations. Defining the power spectrum, $P(k)$ in terms of the
Fourier transform of the density fluctuation:
$P(k)\equiv\left|{\int (\delta\rho/\rho)\exp(ik\cdot x)d^3x}\right|^2$,
one finds that the rms density fluctuations on scales with comoving
wavenumber, $k$ can be written (e.g. Peebles~1980)
\begin{equation}
\left( {\delta\rho\over\rho} \right )^2_k = {k^3\over 2\pi^2}P(k)
\end{equation}

One conventionally considers $P(k) \propto k^n$. This assumes a
scale-free spectrum, namely that there is no preferred scale for
fluctuations set by the underlying physics.  Further, $n=1$ represents a
scale-invariant spectrum as predicted in standard inflation (Steinhardt \&
Turner~1984), up to logarithmic corrections. It has been stressed, coincident
with COBE, that a number of inflationary models which attempt to
circumvent the problems of new inflation also predict measurable
deviations from a scale invariant spectrum, i.e they can predict $n < 1$
(Davis et al.~1992; Gelb et al.~1992; Liddle \& Lyth~1992a,b;
Lidsey \& Coles~1992; Lucchin, Matarrese \& Mollerach~1992; Salopek~1992).
COBE reports a limit on this spectral index of $n = 1.1\pm 0.5$, which is
consistent with scale invariance as well as with significant deviations from
scale invariance. Clearly to get an optimum handle on this spectrum one would
like to compare large-scale and small-scale anisotropies in the CMB.

\vskip .1in

\noindent{\bf{3. CMB anisotropies}}

It is useful to decompose the temperature fluctuations in spherical
harmonics $\Delta T/T(\theta,\phi)\equiv \sum_{lm} a_{lm}Y_{lm}(\theta,\phi)$
and to define the rotationally symmetric quantity
$a_l^2\equiv\sum_m\left|a_{lm}\right|^2$. We choose to consider here the
so-called rms fluctuation, or correlation function at ``zero-lag", as our
contact with experiment. This has a simple interpretation and is sensitive
to a range of physics from both large and small scales.
To be precise we consider
\begin{equation}
\left(  {\Delta T\over T} \right)^2 \equiv C(0)
= {1\over 4\pi}\sum_{l=2}^{\infty} \left[ (a_l^2)_S + (a_l^2)_T \right] W_l,
\label{eqn:rmsdef}
\end{equation}
where the dipole has been subtracted and $W_l$ is a ``window function", which
incorporates the experimental sensitivity to each moment. The window
functions appropriate to the experiments that we will consider in this paper
are  described, and derived, in appendix A. As written, $\Delta T/T$ has
contributions $(a_l)_S^2$ from scalar (density fluctuations) and
$(a_l)_T^2$ from tensor (gravity wave) modes. For the COBE experiment,
the window function is simply due to the gaussian smoothing,
$W_l =  \exp\left[-(l+1/2)^2\sigma^2\right]$ with
$\sigma=0.425\times10^{\circ}$ the smoothing scale (the factor 0.425
relates the gaussian width $\sigma$ to the FWHM quoted by the COBE
group). We will first concentrate on this ``window", where $\Delta T/T$
then represents the rms temperature fluctuation over the observed sky.
Understanding the cosmic variance of the COBE measurement is important
if COBE is to be used to normalize the power spectrum to compare with
other experiments.

Inflation predicts that the $a_{lm}$ are independent gaussian random
variables ($\Rightarrow a_l^2$ is $\chi^2$-distributed with $2l+1$ d.o.f.)
so the rms fluctuation as defined is itself a stochastic quantity (Abbott
\& Wise~1984; Krauss \& White~1992; White~1992). It might be thought
that measurements made over the entire sky afford the opportunity to
make independent measurements of the quantities which have a cosmic
variance, and so reduce it. However from the above, it is clear that this
cannot be the case.
No measurement of $\Delta  T/T$ on our sky, no matter how accurate nor how
many independent  measurements it corresponds to, can reduce the cosmic
variance for a given $l$.
Determining each $a_l^2$ requires an all-sky average (we will not address the
important issue of the effects of incomplete sky coverage in this paper as we
are interested here in the {\em irremovable} cosmic variance.
The issue of finite sky coverage is addressed in Scott et al.~1993.).
The error in measuring $a_l^2$ will be reduced by more sky measurements, but
not the uncertainty in going from $a_l^2$ to the average over an ensemble of
universes $\langle a_l^2\rangle_{ens}$.
(However, as we will see, window functions relevant to small scale
measurements can  give dominant contributions to $\Delta T/T$ from
higher modes, which  have more degrees of freedom and smaller cosmic
variance, so that the  effect of this uncertainty can in principle be
lessened, provided one has adequate sky coverage.)

It is not possible in general to find a closed form analytic expression
for the  distribution of $\Delta T/T$.  We outline a simple and accurate
numerical procedure for obtaining the  distribution in appendix B.
It is straightforward, however, to calculate the  first few moments of the
distribution (since $\Delta T/T$ is a sum of {\em independent},
$\chi^2$-distributed terms) which give us an estimate of the size of
the variance.  Write
\begin{equation}
\left(  {\Delta T\over T} \right)^2   =
\sum_{l} {a_l^2\over\langle a_l^2\rangle} (l+1/2) \Delta_l
\ \ \ \ \ \mbox{with}
\ \ \ \ \ \Delta_l={2\over 2l+1}{\langle a_l^2\rangle\over 4\pi}\ W_l
\label{eqn:deldef}
\end{equation}
so $\overline{(\Delta  T/T)^2}\equiv\langle(\Delta T/T)^2\rangle=
\sum_l (l+1/2)\Delta_l$.
If we define
$\Delta^{(n)}=\langle((\Delta T/T)^2-\overline{(\Delta T/T)^2})^n\rangle$
then we can use $\Delta^{(2)}$ as an estimate of the (cosmic) variance of
$\Delta T/T$ and $\Delta^{(3)}$ as an indicator of the asymmetry of the
distribution about the mean (and departures from normality). It is
straightforward to compute\footnote{Our result for $\Delta^{(2)}$ differs
from Bond \& Efstathiou~(1987) by a factor of 2.}
\begin{equation}
\Delta^{(2)} =  \sum_l (l+1/2)\Delta_l^2
\ \ \ \ \ \ \ \ \    \Delta^{(3)} = 2\sum_l (l+1/2)\Delta_l^3.
\label{eqn:moments}
\end{equation}

For $n=1$ Sachs-Wolfe fluctuations ($\langle a_l^2\rangle\sim (l+1/2)^{-1}$,
$l\gg 1$), the rms fluctuation
$\overline{(\Delta T/T(\sigma))^2}\sim\log\sigma$
while the higher moments are all finite\footnote{In practice of course the rms
fluctuation will not diverge as $\sigma\rightarrow 0$.  Physics associated
with the size and width of  the last scattering surface damps the $a_l^2$ for
high $l$. This can also  be aided by spectral tilt.}. Thus the ``cosmic
variance" becomes less important at smaller scales.  It is at these small
angular scales (less than about $2^{\circ}$) however that the theoretical
uncertainties due to cosmological model dependence and non-linear effects
become progressively more severe.

To better illustrate the form of the distribution for $\Delta T/T$ we have
compared two approximations to the exact distribution (which we checked
by doing a Monte-Carlo analysis in which the $a_l^2$ were drawn at random
from $\chi_{2l+1}^2$ distributions and $(\Delta T/T)^2$ computed using
equation \ref{eqn:rmsdef}.). In the case of only a scalar component in
fluctuations of the $n=1$ Sachs-Wolfe form (specifically
$\langle a_l^2\rangle_S= (l+1/2)^{-1}$ for $l\ge 2$), we show in Figure 1a
the distribution of $\Delta T/T$ smoothed to $10^{\circ}$ obtained from the
method outlined in appendix B and the Monte-Carlo calculation (see also
Table \ref{tab:som}).  These are in good agreement. Also shown
are fits obtained from matching the mean and variance of the distribution
to a $\chi^2$ form ($P(x)\propto x^{c_1}e^{-c_2x}$) and a gaussian. As can
be seen, it is not well fit by a gaussian. The fit to a single $\chi^2$ form
gets progressively better for larger angles, or tilted spectra. This is
because the sum is dominated by the lower multipoles in these cases and
is therefore closest to being represented by a single $\chi^2$ distribution.
It is important to note that if one attempts to fix the normalization of the
power spectra by fitting $\Delta T/T$ to the COBE measurement, there is
an uncertainty due to cosmic variance of $\approx 20\%$ (figure 1a).  This
normalization uncertainty will affect any comparison with smaller scale
experiments and reduce what can be inferred.  In the future it may be better
to normalize the spectrum not by the COBE $\Delta T/T$ but by something
less sensitive to the low $l$ multipoles.  We will return to this point later.

For smoothing angles smaller than $10^{\circ}$, one can use the second moment
of the distribution as an equivalent gaussian error for the purposes of
computing confidence levels.  We again stress that the distribution is
{\em not gaussian} but for quantities depending only on the asymptotic form
of the distribution (e.g.  90, 95, 99\% confidence regions) such an
approximation is adequate. Nevertheless this skewness should not be ignored
when utilizing measurements of either $\Delta T/T$ or the correlation function
$C(\theta)$ to place limits on microphysical parameters such as the scale of
inflation.

\vskip .1in

\noindent{\bf{4. The Anisotropy at $1^{\circ}$ Scales}}

On degree scales, for the scalar fluctuations, one must consider in
addition to the Sachs-Wolfe terms, other terms such as the Doppler
contribution (from motions on the last scattering surface). These
contributions depend on the cosmological model (e.g. the Hubble constant,
$\Omega_B$ and the re-ionization history of the universe), which
governs the growth of fluctuations inside the horizon, and generally
require numerical solution of the Boltzmann equation for the evolving
photon distribution for an accurate estimation.
To calculate $\langle a_l^2\rangle$ we have used the radiation ``power
spectrum" $k^3 W_T^2(k)$ of Vittorio \& Silk~(1992), which includes a
``transfer function" for the evolution of modes which have crossed the
horizon.
As an illustrative example of the non-Sachs-Wolfe terms, we calculated the
$\langle a_l^2\rangle$ for a CDM model with $\Omega=1, h=0.5$ and
$\Omega_B=0.03,0.1$ (consistent with the range favoured by nucleosynthesis:
Smith et al.~1992, Krauss \& Romanelli~1990, Walker et al.~1991).
For $n\ne 1$ we multiplied the Sachs-Wolfe result by the ratio of
the full calculation to Sachs-Wolfe for $n=1$.  This approximation is
expected to be good to better than 10\%, as can be seen from analytic
work (Atrio-Barandela \& Doroshkevich~1993).
We normalized the $a_l^2$ to fit the COBE central value at $10^{\circ}$.
The power spectra are shown in Figure 2 for $n=1.0$ for the range of
$k\equiv lH_0/2$ of relevance here.
(An unprocessed Sachs-Wolfe power spectrum would go as $k^{n-1}$.)
The $\langle a_l^2\rangle$ are an integral of the ``power  spectrum" with
measure $j_l^2(2k/H_0)dk/k$ --- in effect an average of the ``power spectrum"
around $k\approx lH_0/2$.
Hence the `Doppler  enhancement' of the power spectrum on degree scales
translates to an increase in the $\langle a_l^2\rangle$ for large $l$ over
the Sachs-Wolfe result.
To obtain $\langle(\Delta T/T)^2\rangle$, one performs a weighted sum of
$\langle a_l^2\rangle$ with weight $W_l$ as in equation (\ref{eqn:rmsdef}).
Plotted in Figure 2 are the window functions for the COBE, MIT,
UCSB South Pole (SP91) and MAX experiments to show the range
of multipoles probed by each\footnote{Recall in our convention the window
function is convolved with $a_l^2$ which fall with $l$ in the Sachs-Wolfe
region.}.

In Figure 3 we display the predicted value of $\Delta T/T$ (for
$\Omega_B=0.03,0.10$) for SP91 vs $n$, for a spectrum which is normalized
to fit the COBE measured $\Delta T/T$.
(For a more detailed comparison of SP91 with COBE in the context of
standard CDM see Gorski, Stompor \& Juszkiewicz~1992).
The solid lines represent the 90\% confidence level predictions assuming a
tensor component is present in the amount predicted by power-law inflation
(Davis et al.~1993) and the dashed lines assume that the tensor component is
absent (see section 5).
Note that the entire uncertainty in the values shown comes from cosmic
variance of the COBE result, with present observational uncertainty ignored
and no uncertainty for SP91.
This curve allows a determination of the constraints on $n$ which may be
possible from a comparison of SP91 and COBE. Concentrating for now on the
solid  curves, giving the full result (including tensor contributions to be
described next), we see that even an exact measurement of $\Delta T/T$
by SP91 may only constrain $n$ to within $\approx\pm 0.2$, due to the
COBE cosmic variance.  Including a SP91 measurement uncertainty might,
depending on its value and the measured $\Delta T/T$, allow the whole
range $n =0.5-1$. The corresponding predictions for the MAX experiment are
shown in Figure 4.

In principle one might reduce the uncertainty on the normalization of the
$\langle a_l^2\rangle$ inferred from COBE (and hence the theoretical
uncertainty on smaller scales) by using not just $\Delta T/T$ but the
whole correlation function $C(\theta)$, which contains more information.
Because $C(\theta)\propto\sum_l a_l^2 P_l(\cos\theta)W_l$ probes
a different linear combination of $a_l^2$ at each $\theta$, one might hope
for increased sensitivity to higher $l$ modes which have smaller cosmic
variance.  Nevertheless, in practice the present sensitivity is such that
most of the signal comes from small $l$, so that any treatment of the
existing data cannot decrease our estimate of the errors to the point where
cosmic variance is irrelevant.
The relative uncertainty and size of the $\langle a_l^2\rangle$ are shown
in Table \ref{tab:som} for reference.

As Figure 2 shows, the MIT experiment probes a larger range of $l$ than
COBE and the effects of cosmic variance on the normalization inferred
from this experiment are less.  We find that the normalization uncertainty
is $\sim 13\%$ (thus the MIT experiment would be more useful than COBE
for fixing the normalization of the spectrum if this experiment were to have
extensive sky coverage and reduced experimental errors).
As shown in Figure 1b $(\Delta T/T)^2$ still has significant skewness and
is  not well approximated by a gaussian of mean
$\langle (\Delta  T/T)^2\rangle$. Again the correlation function
(assuming small errors at  each angle) would contain more information
than $\Delta T/T$, but would  be more difficult to analyze.
Even so it would be useful to have a measurement of the anisotropy on scales
large enough that sensitivities to the doppler peak ($\Omega_B$ and amount of
tensor component) and re-ionization are small, while being such that low $l$
modes to not contribute significantly e.g. a ``full sky MIT" correlation
function or an experiment intermediate between Tenerife (Davies et al.~1987),
which measures fluctuations on $8^{\circ}$ scales, and SP91 on degree scales.

\vskip .1in

\noindent{\bf{5. Tensor versus Scalar modes}}

It was pointed out in Krauss \& White~(1992) that gravitational wave (i.e
tensor modes) could play an important role in interpreting the COBE
anisotropy if it stems from inflation, so that these should not be ignored.
As a result, several groups examined the ratio of scalar to tensor modes
predicted to result from various inflationary models (Davis et al.~1992;
Dolgov \& Silk~1993; Krauss~1992; Salopek~1992). In Davis et al.~(1992),
a general connection was pointed out between the scalar to tensor
ratio predicted by inflation and the spectral tilt as given by $n\ne 1$.
For power law inflation this relation takes the approximate form
$n\approx 1-1/7(T/S)$ (Davis et al.~1992) where $(T/S)$ represents the
ratio of the {\em expectation values} of the tensor and scalar quadrupole
moments.
As a result it appears that this ratio, if it were measurable, could provide
important information about an inflationary epoch in the early universe
(Davis et al.~1992). It is important to note however that while the ratio of
the expectation values are fixed by the equation of state during inflation,
the separate cosmic variance can alter the actual ``observed" ratio of
quadrupoles significantly from that predicted by $T/S$.
For two $\chi^2$-distributed random variables with $2l+1$ d.o.f., the ratio
has the distribution
\begin{equation}
P(R)dR = {\Gamma(2l+1)\over\Gamma^2(l+1/2)}{R^{l-1/2}\over(1+R)^{2l+1}}
\ \ \ R={a_l^2/a_l'^2\over \langle a_l^2\rangle/\langle a_l'^2\rangle}
\end{equation}
In Figure 5 we have plotted $P(R)$ and the integral $\int_0^R P(r)dr$ for
the quadrupole ($l=2$).  Note that the distribution is peaked at a value
below 1 but has a significant large $R$ tail.
With $P(R_1<R<R_2)=95\%$ we find that $R$ can range from $0.2-5$!  Hence
fluctuations with $(a_2^2)_T/(a_2^2)_S$ 5 times (or 1/5) the ratio
$\langle a_2^2\rangle_T/\langle a_2^2\rangle_S$ are not improbable.
Thus we see that a direct measurement of the quadrupole ratio
$(a_2^2)_T/(a_2^2)_S$ could not provide much information on the actual
equation of state during inflation!
One needs to look at quantities (like $\Delta T/T$) that have smaller cosmic
variance such as the power spectrum normalized quadrupoles $Q_{rms-PS}$
(Smoot et al.~1992).  In this case the conversion to $T/S$ is much less
uncertain. The term quadrupole in this context should therefore be interpreted
with care.

Nevertheless any confirmation of a tensor component to the CMB
anisotropy on COBE scales or below would be of great significance for the
interpretation of the COBE measurement, whether or not it pins down models.
However, our results point out the difficulty in making such a
confirmation based on comparing CMB anisotropies on $1^{\circ}$ scales
with the COBE result.  The solid curve in Figure 3 includes the tensor
contribution in the amount prescribed by power law inflation as a function
of $n$. If the tensor contribution were absent for some reason (dashed curve),
the predicted $\Delta T/T$ at small angles would be increased.  This is
because the $(a_l^2)_T$ begin to damp (to be conservative) at
$l\simeq\sqrt{1+z_{dec}}\approx 30$ due to redshifting of the gravity waves
(Starobinsky~1985; Turner et al.~1993).
It is therefore smaller than the scalar for high $l$ modes, so that its
relative contribution to COBE is larger than it is to small angle experiments
[Note that the spectral tilt required to have a large tensor component
increases the contribution of low $l$ modes to the sum, reducing the
sensitivity to the Doppler peak and makes tensor and scalar contributions
similar.]
While this provides some sensitivity to a tensor component in principle,
for SP91 the maximum variation is only slightly greater than the
(90\% confidence level) irremovable uncertainty due to cosmic variance.
Including the uncertainty in $\Omega_B$ (and $n$) it seems that a comparison
of $1^{\circ}$ scale measurements with COBE can at best marginally probe
the expected tensor contribution to $\Delta T/T$.
The situation on still smaller angular scales is less pessimistic in this
regard, but one should remember that these scales are most sensitive to
non-linear effects and assumptions about the cosmological model, as shown
in Figure 4.

The results of this section, and those of section 4, demonstrate the
essential problem involved in comparing the COBE results to those of
small angular scale experiments.  If the window function of such experiments
probes different $l$ than COBE, the $20\%$ COBE cosmic variance can blur the
comparison.
To overcome this normalization uncertainty, we suggest using the COBE
correlation function or $\Delta T/T$ as measured by the MIT experiment to
fix $\langle a_2^2\rangle$.  In the latter the window functions for small
scale experiments can overlap the MIT window function significantly, and
correlations between the variance affecting the normalization and the variance
in the small angular scale predictions must be taken into account.

\vspace{.1in}

\noindent {\bf{6. Cosmological Model Dependence}}

As alluded to earlier, while the COBE measurement probes scales outside
the horizon today and is mostly insensitive to details of the cosmological
model, on smaller scales the predictions are model-dependent.
Apart from uncertainties in cosmological parameters (such as $H_0$ and
$\Omega_B$) which induce uncertainties in the prediction of $\Delta T/T$,
the major unknown arises from the possiblity of early re-ionization and
re-scattering of the primoridial CMB fluctuations.
The unbiased CDM model, precisely that normalized to COBE, allows adequate
non-linearity at early epochs that re-ionization and generation of a secondary
``last scattering surface" (LSS) is not implausible (Tegmark \& Silk~1993).
The ionization of hydrogen imposes such modest energetic requirements that
rare fluctuations of low mass can provide a plausible source of ionizing
photons or particles (an effect that must have occurred by $z=5$ to satisfy
the Gunn-Peterson constraint).
Such re-ionization would move the LSS in to lower redshift and reprocess
fluctuations on scales smaller than the horizon size at the new LSS.
Indeed for a range of (high) CDM normalizations with $n=1$ one can reduce
degree scale fluctuations by a factor $\sim 2-3$.  However if the fluctuations
are small (corresponding to biassing factors $b\approx 2$), the spectrum is
tilted or is that of a mixed dark matter model, it appears that degree scale
fluctuations remain largely unaffected.

Secondary fluctuations induced on very small (arc-minute) scales and their
detection could eventually provide a useful constraint on the last scattering
surface.
Until then however primary fluctuations on degree scales are potentially
more sensitive to the last scattering surface (re-ionization) than to the
possible presence of tensor modes.

\vspace{.1in}

\noindent {\bf{7. Conclusions}}

After the initial flurry of excitement over the COBE data and its potential
ability --- in combination with other CMB measurements --- to probe the
details of models for the generation of primordial density fluctuations, it
is time to realistically assess the practical and theoretical limitations
inherent in any such procedure.
This is  particularly important as new measurements of CMB anisotropies
on $1^{\circ}$ scales are expected.  Our results indicate that the statistics
of CMB anisotropies predicted by inflation play a fundamental and  irremovable
role in limiting the information we can gain from the data.  Cosmic variance,
even with an exact COBE result, still provides  uncertainty in the extracted
spectral index $n$.
By incorporating this variance, we have been able to quantitatively
display how limits on $\Delta T/T$ at small angles can translate to
limits on $n$.  Going to smaller scales, such as those comparable to
that of observed large scale structures, would provide a longer lever
arm and in principle tighter constraints on $n$.  However the theoretical
uncertainties due to causal  effects and their relation to the growth of
fluctuations are much larger in this case.  Finally, while observation of a
tensor component to the CMB  anisotropy would be of great interest, we have
shown that the predicted levels are not easily extractible from the data.

To reduce the effects of cosmic variance we suggest that the full correlation
function of COBE or the MIT experiment would give a theoretically
cleaner normalization of the power spectrum than the COBE $\Delta T/T$.
This can then be used with the South Pole or MAX experiments to constrain
models of structure formation or cosmology.  The normalization uncertainty
in this case can be reduced by perhaps a factor of 2. However this would still
make a detection of a gravitational wave signal very difficult, coupled as it
is with the uncertainties in $n$, $\Omega_B$ and re-ionization history.

Inflation is an attractive theory in part because it makes unambiguous
predictions which can be tested.  Similarly, measurements of the CMB
anisotropy still  provide the best hope we have of extracting information
on the primordial power spectrum of density perturbations.  As we have
shown however, definitive answers will require careful attention to the
theoretical uncertainties. Inflation predicts probability distributions
for observed CMB anisotropies. It will be important to determine what
CMB measurements can rule out conclusively. On the other hand, it is
equally important not to overstate the case.

\vspace {.1in}


\noindent{\bf Acknowledgements}

The authors would like to thank N. Vittorio for providing power spectra and
R. Davis, J. Jubas, C. Lineweaver, P. Richards, J. Schuster, G. Smoot and
M. Wise for useful discussions.
Also D. Scott for comments on the manuscript, M. Turner for illuminating
discussions on the damping of tensor induced anisotropies and S. Dodelson
for discussions of window functions and for kindly providing his code for
evaluating spherical harmonics.

This work was supported by grants from the DOE, NSF and TNRLC.

\clearpage

\noindent {\bf Appendix A}

The sensitivity of experiments to the spherical harmonic decomposition of the
CMB temperature fluctuations is conventionally described in terms of window
functions $W_l$ as in equation (\ref{eqn:rmsdef}).  For completeness, and
to fix our conventions, we describe in this appendix the window functions
appropriate to the COBE, MIT, UCSB South Pole and MAX experiments.
The functions are shown in Figure 2.

\vskip .1in

\noindent {\bf COBE and MIT}

A simple way to view the window functions is in terms of a mapping
$Y_{lm}\rightarrow\tau_{lm}Y_{lm}$ (no sum) describing the weighting the
experiment gives to each mode. An example makes this clear.
The most straightforward such mapping is that caused by a gaussian smoothing
function of width $\sigma\ll 1$ (which could be due to antenna response or
applied to the data ``by hand")
$$ Y_{lm}(\theta_0,\phi_0) \rightarrow
\int {\theta d\theta d\phi\over 2\pi\sigma^2} \exp\left[
{-\theta^2\over 2\sigma^2} \right] Y_{lm}(\Omega_0+\Omega) $$
$$\approx\exp\left[{-(l+1/2)^2\sigma^2\over 2}\right]Y_{lm}(\theta_0,\phi_0)$$
(for details see White, 1992).
If we substitute $\delta T/T=\sum_{lm} a_{lm} \tau_{lm} Y_{lm}$ into
$$ \left( {\Delta T\over T} \right)^2 \equiv
 \left\langle\left( {\delta T\over T} \right)^2 \right\rangle_{sky} =
\sum_l {a_l^2\over 4\pi} W_l $$
where $\langle\cdots\rangle$ indicates an average over the whole sky, and
use the orthonormality of the $Y_{lm}$ we obtain the window for the COBE
experiment (Smoot et al.~1992) $W_l=\exp[-(l+1/2)^2\sigma^2]$ shown
in Figure 2.
In this case $\Delta T/T$ corresponds to an rms temperature fluctuation over
the observed sky, smoothed on a scale $\sigma=0.425\times 10^{\circ}$.
(The COBE maps with no additional smoothing correspond to
$\sigma=0.425\times 7^{\circ}$, not shown in Figure 2).
We find the same factor appearing in the other window functions due to the
finite resolution of the antenna (which we assume to be gaussian).  In what
follows we will make the approximation of a ``perfect" antenna, knowing we can
put the antenna response in at the end.

The MIT experiment (Page et al.~1990) also calculates $\Delta T/T$ from a
map of the sky and so the form of the window is identical to that of COBE.
The smoothing scale is $\sigma = 0.425\times 3.8^{\circ}$.

\vskip .1in

\noindent {\bf South Pole and MAX}

Now consider the UCSB South Pole experiment (Gaier et al.~1991) and focus on
the measurement of the so-called $(\Delta T/T)_{rms}$ on one patch of the sky,
centered at $\theta_0,\phi_0$. (Since it is our purpose here to show the effect
of cosmic variance on inferences made from this experiment, and not to fit the
data, we will not be concerned with multiple, correlated patches of the sky.
For a discussion of the fitting of the data see Bond et al.~(1991);
Dodelson \& Jubas~(1992).  Note that in Bond et al.~(1991) a factor of
$4\pi/(2l+1)$ is missing from equation (1).)
The temperature difference at $\theta_0,\phi_0$ is measured by ``chopping" the
beam at fixed $\theta=\theta_0$ through
$\phi(t)=\phi_0+\alpha\sin(2\pi\nu t)$ and sampling $\delta T/T(\theta,\phi)$
with weight sign$(\phi(t)-\phi_0)$.
Recalling the temperature difference is twice the weighted average,
the mapping we want is then (see Dodelson \& Jubas~1992)
$$ Y_{lm}(\theta_0,\phi_0) \rightarrow
2\nu \int_0^{1/\nu} dt\ \mbox{sign}(\phi(t)-\phi_0) Y_{lm}(\theta_,\phi(t)) $$
$$ = 2i H_0(m\alpha) Y_{lm}(\theta_0,\phi_0) $$
where in the last line we have written the integral $dt$ as an integral
$d(\phi(t)-\phi_0)$, extracted the $\exp[im\phi]$ dependence from
$Y_{lm}$ and used the identity (Gradshteyn \& Ryzhik~1980)
$$ H_n(x) = {2(x/2)^{n}\over\sqrt{\pi}\,\Gamma(n+1/2)}
   \int_0^1 dt\ (1-t^2)^{n-1/2} \sin(xt) $$
for the Struve function.
At this point we could proceed to define the sky rms fluctuation by averaging
over $\theta_0,\phi_0$ as before (now with an $m$ dependent $\tau_{lm}$).
However the ``rms" $\Delta T/T$ quoted by the SP group does not come from
this procedure.  Instead we now argue that the small angular scales probed
by this experiment make it sensitive only to high $l$ multipoles.  These
multipoles have a small cosmic variance, thus our universe at $\theta_0,\phi_0$
should be similar to the ensemble average of all universes at the same point.
We can then define
$(\Delta T/T)^2\equiv\langle (\delta T/T)^2\rangle_{ens}$
and use (dropping the subscript $ens$)
$$ \langle a_{l'm'}a_{lm}^{*}\rangle= {\langle a_l^2\rangle\over (2l+1)}
\delta_{l'l}\delta_{m'm} $$
to write
$(\Delta T/T)^2=\sum_{lm}\langle a_l^2\rangle |\tau_{lm}Y_{lm}|^2/(2l+1)$.
Reinstating the finite beam width the window so defined is then
$$ W_l = \sum_{m=-l}^l W_{lm} =
{4\pi\over (2l+1)} \exp\left[-(l+1/2)^2\sigma^2\right]
\sum_{m=-l}^l \left| Y_{lm}(\theta_0,\phi_0)\right|^2 4H_0^2(m\alpha) $$
For the configuration of (Gaier et al.~1991) $\theta_0=27.75^{\circ}$,
$\alpha\sin\theta_0=1.5^{\circ}$ and $\sigma=0.425\times 1.65^{\circ}$.

If the effects of sky rotation and smooth vs. step scanning are ignored
the form of the window function for the MAX experiment is very similar to
SP91.  However the MAX experiment uses a phase lock-in which extracts the
frequency component of the sky signal which matches the chop frequency $\nu$.
Extracting only this component corresponds to replacing the weight
sign$(\phi(t)-\phi_0)$ with $(\phi(t)-\phi_0)/\alpha$ in the $Y_{lm}$ mapping
defined above.  The remaining integral over $d\phi$ is then not $H_0$ but a
Bessel function: $\pm(\pi/2)J_1(|m\alpha|)$.  The window function is as above
for SP91 but with $2H_0\rightarrow\pi J_1$.
[This explains why our window function and predictions for MAX differ from
that of other authors.]
The parameters (for the $\mu$-Pegasii scan, Meinhold et al.~1992) are
$\theta_0=90^{\circ}, \alpha=0.65^{\circ}, \sigma=0.425\times 0.5^{\circ}$.

We can test our assumption that the cosmic variance is ``small".
Recalling that the $a_{lm}$ are gaussian independent complex variables
it is straightforward to show that for an experiment covering the whole
sky\footnote{At the other extreme if the experiment looks at only one point
on the sky the variance is twice the square of the mean, as expected for
(the square of) a gaussian random variable.  The intermediate case is
discussed in Scott et al.~1993.}:
$$ \left\langle \left({\Delta T\over T}\right)^2 \right\rangle =
  \sum_{lm} {\langle a_l^2\rangle\over 4\pi} W_{lm} $$

$$ \left\langle\left[\left({\Delta T\over T}\right)^2\right]^2\right\rangle -
 \left\langle \left({\Delta T\over T}\right)^2\right\rangle^2 =
  2\sum_{lm} \left( {\langle a_l^2\rangle\over 4\pi} W_{lm}\right)^2 $$
where the $Y_{lm}$ factor is to be removed from $W_{lm}$.
Notice that these reduce to equation (\ref{eqn:moments}) when $W_{lm}$ is
$m$-independent.
Keeping in mind that the distribution of $\Delta T/T$ will not in general be
gaussian, this allows us to derive an estimate for the size of the cosmic
variance associated with each experiment.
We find the uncertainty (at 1``$\sigma$") for an $n=1$ spectrum to be
4 and 1\% for the SP91 and MAX windows respectively.
This would be larger for $n<1$.

\vskip .2in

\noindent {\bf Appendix B}

The probability distribution for the correlation function predicted by
cosmic variance is not well fit by a gaussian or $\chi^2$ form.  In this
appendix we derive a simple expression for the distribution which is amenable
to numerical computation.  We hope that this will facilitate comparisons of
theory with data when they become available.

For now let us consider the correlation function predicted by a one component
spectrum (either tensor or scalar).  We will keep the $\theta$ dependence here
for generality, and defer consideration of two component spectra till later.
For COBE and MIT, by definition
$$ C(\theta) \equiv {1\over 4\pi} \sum_{l=2}^{\infty} a_l^2
  P_l(\cos\theta) W_l  =  \sum_{l=2}^{\infty} c_l y_l $$
where we have defined $y_l=a_l^2/\sigma_l^2$ with
$\sigma_l^2=\langle a_l^2\rangle/(l+1/2)$ and
$c_l=(\sigma_l^2/4\pi)P_l(\cos\theta)W_l$.
(Note $c_l$ reduces to $\Delta_l$ of equation (\ref{eqn:deldef}) for
$\theta=0$)
For the Sachs-Wolfe part of the spectrum the $c_l\sim l^{-2}W_l$
and so fall off rapidly with $l$.
Each $y_l$ is an independent random variable  with distribution
$$P_{\chi}^{(l)}(y) = {y^{l-1/2} e^{-y}\over \Gamma(l+1/2)} \theta(y)$$
with $\theta$ the Heaviside (or step) function.
The correlation function is thus a random variable with distribution
given by the convolution
$$ P(C) = \lim_{N\rightarrow\infty}
   \left( \prod_{l=2}^N c_l^{-1} \right)
   P_2\circ P_3\circ\cdots\circ P_N(C) $$
where $P_l(x)\equiv P_{\chi}^{(l)}(x/c_l)$.  To simplify this we take the
Fourier transform (compare with Cayon et al.~1991)
$$ {\cal F}[P]  =  \lim_{N\rightarrow\infty}
  \prod_{l=2}^N { {\cal F}[P_l] \over c_l}
  = \prod_{l=2}^{\infty} \left( 1-i\omega c_l\right)^{-(l+1/2)} $$
This can be easily inverted.  Writing
$\log\left({\cal F}[P]\right)=-a_{\omega}+ib_{\omega}$ we need to
calculate
$$ P(C) = {1\over 2\pi} \int_{-\infty}^{\infty} d\omega\ e^{-a_{\omega}}
   \cos\left( b_{\omega} - \omega C \right) $$
where
$$ a_{\omega}  \equiv  \sum_{l=2}^{\infty} (l+1/2)
    \log\sqrt{ 1+(\omega c_l)^2 }  $$
and
$$ b_{\omega}  \equiv  \sum_{l=2}^{\infty} (l+1/2)
    \arctan(\omega c_l) $$
This expression is exact and the integral can be efficiently done numerically.
One only needs to find $a_{\omega}$ and $b_{\omega}$ once for each set of $c_l$
to evaluate the distribution for all $C$.
The results shown in Figure 1 come from taking $\theta=0$ and
$\langle a_l^2\rangle=(l+1/2)^{-1}$.
To include additional (independent) components of the spectrum one just has
$a_{\omega}=\sum_i a_{\omega}^{(i)}$ and $b_{\omega}=\sum_i b_{\omega}^{(i)}$
where $a^{(i)}, b^{(i)}$ are computed from $c_l^{(i)}$ as above.

\clearpage

\begin{table}[h]
\begin{center}
\begin{tabular}{|c|c|c|c|c|} \hline
$l$ & \% Uncertainty & $10^{\circ}$ & $7^{\circ}$ & $3.8^{\circ}$ \\ \hline
   2 & 63 & 1.00 & 1.00 & 1.00 \\
   3 & 53 & 0.69 & 0.70 & 0.71 \\
   4 & 47 & 0.51 & 0.53 & 0.55 \\
   5 & 43 & 0.40 & 0.43 & 0.45 \\
   6 & 39 & 0.32 & 0.35 & 0.37 \\
   7 & 37 & 0.25 & 0.29 & 0.32 \\
   8 & 34 & 0.20 & 0.25 & 0.28 \\
   9 & 32 & 0.17 & 0.21 & 0.25 \\
  10 & 31 & 0.13 & 0.18 & 0.22 \\
  11 & 29 & 0.11 & 0.15 & 0.20 \\
  12 & 28 & 0.09 & 0.13 & 0.18 \\
  13 & 27 & 0.07 & 0.12 & 0.16 \\
  14 & 26 & 0.06 & 0.10 & 0.15 \\
  15 & 25 & 0.04 & 0.09 & 0.13 \\ \hline
$\sum_{l=2}^{\infty} $ & --- & 4.19 & 5.01 & 6.48 \\
\hline
\end{tabular}
\end{center}
\caption{The relative uncertainty and expected size of the multipoles measured
by the correlation function from COBE (smoothed on $10^{\circ}$ and $7^{\circ}$
scales) and from MIT (smoothed on $3.8^{\circ}$ scales) for an $n=1$
Sachs-Wolfe spectrum. For $n<1$ the multipoles drop off faster with $l$.
The quadrupoles are normalized to 1 and the sum $l=2,\cdots,\infty$ is listed
below each column.  The uncertainty shown is the second moment of the
$\chi^2_{2l+1}$-distribution for $a_l^2$.}
\label{tab:som}
\end{table}

\clearpage

{\bf References}

\begin{itemize}
\item[] Abbott, L.F. \& Wise, M.B., 1984, ApJ 282, L47
\item[] Atrio-Barandela, F. \& Doroshkevich, A.G., 1993, preprint CfPA-TH-93
\item[] Bertschinger, E. \& Dekel, A., 1989, ApJ 336, L5
\item[] Bertschinger, E., Gorski, K.M., \& Dekel, A., 1990, Nature 345, 507
\item[] Bond, J.R. \& Efstathiou, G., 1987, MNRAS 226, 655
\item[] Bond, J.R., et al.~1991, Phys. Rev. Lett. 66, 2179
\item[] Cayon, L., Martinez-Gonzalez, E. \& Sanz, J.L., 1991, MNRAS 253, 599
\item[] Davies, R.D., et al.~1987, Nature 326, 462
\item[] Davis, R.L., et al.~1992,  Phys. Rev. Lett. 69, 1856
\item[] Dodelson, S. \& Jubas, J.M., 1992, Phys. Rev. Lett. 70, 2224
\item[] Dolgov, A. \& Silk, J., 1992, Phys. Rev. D47, 2619
\item[] Gaier, T., et al.~1992, ApJ 398, L1
\item[] Gradshteyn, I.S. \& Ryzhik, I.M., 1980, ``Tables of integrals, series
and products", 4th ed. (Academic, New York)
\item[] Gelb, J.M., Gradwohl, B.A. \& Frieman, J.A., 1993, ApJ 403, L5
\item[] Gorski, K.M., Stompor, R. \& Juszkiewicz, R., 1992, ApJ 410, L1
\item[] Krauss, L.M., 1992, preprint YCTP-P21-92
\item[] Krauss, L.M. \& Romanelli, P., 1990, ApJ 358, 47
\item[] Krauss, L.M. \& White, M., 1992, Phys. Rev. Lett. 69, 869
\item[] Liddle, A.R. \& Lyth, D.H., 1992a, Phys. Lett. B291, 391
\item[] Liddle, A.R. \& Lyth, D.H., 1992b, Phys. Lett. B279, 244
\item[] Lidsey, J.E. \& Coles, P., 1992, MNRAS 258, 57
\item[] Lucchin, F., Matarrese, S. \& Mollerach, S., 1992, ApJ 401, L49
\item[] Meinhold, P., et al.~1992, ApJ 406, 12
\item[] Page, L.A., Cheng, E.S. \& Meyer, S.S., 1990, ApJ 355, L1
\item[] Peebles, P.J.E., 1980, ``The large scale structure of the universe",
(Princeton U.P., New Jersey)
\item[] Salopek, D.S., 1992, Phys. Rev. Lett. 69, 3602
\item[] Scott, D., Srednicki, M. \& White, M., 1993, submitted to ApJ Lett.
\item[] Smith, M., Kawano, L.H. \& Malaney, R.A., 1992, ApJ(Supp) 85, 219
\item[] Smoot, G.F., et al.~1992, ApJ 396, L1
\item[] Starobinsky, A.A., 1985, Sov. Astron. Lett. 11, 133
\item[] Steinhardt, P.J. \& Turner, M.S., 1984, Phys. Rev. D29, 2162
\item[] Tegmark, M. \& Silk, J., 1993, submitted to ApJ
\item[] Turner, M.S., White, M. \& Lidsey, J.E., 1993, Phys. Rev. D, submitted
\item[] Vittorio, N. \& Silk, J., 1992, ApJ 385, L9
\item[] Walker, T.P., et al.~1991, ApJ 376, 51
\item[] White, M., 1992, Phys. Rev. D46, 4198
\end{itemize}

\clearpage

\centerline{ {\bf Figure Captions} }

\begin{itemize}
\item[Figure 1] Results of a Monte-Carlo calculation of $\Delta T/T$ for the
(a) COBE and (b) MIT experiments assuming $\langle a_l^2\rangle_S=(l+1/2)^{-1}$
and $\langle a_l^2\rangle_T=0$.  Also shown are the results obtained using the
method of appendix B and fits to a $\chi^2$ form (dotted line) and a gaussian
(dashed line) with the same mean and variance as the Monte-Carlo distribution.
\item[Figure 2] Power spectra $k^3 W_T^2(k)$ for a CDM model with
$\Omega=1, h=0.5, n=1$ and $\Omega_B=0.03$ (solid line) and
$0.10$ (dotted line) as a function of $l\equiv 2k/H_0$.
Also shown are the window functions $W_l$ (as described in appendix A) versus
multipole number $l$ for COBE, MIT, SP91 and MAX.
\item[Figure 3] Predictions for SP91, assuming the COBE central value of
$\Delta T/T$, as a function of $n$.
Error bars are the COBE cosmic variance induced 90\% confidence level
uncertainties, neglecting experimental errors and the cosmic variance
associated with SP91.
Points are given for both scalar and tensor contributions (solid) and scalar
only (dashed).  Each group of points is at the same $n$, the offset is for
ease of viewing only.  Higher points represent $\Omega_B=0.10$ and low
points $\Omega_B=0.03$.
\item[Figure 4] Predictions for MAX, assuming the COBE central value of
$\Delta T/T$, as a function of $n$.
Error bars are the COBE cosmic variance induced 90\% confidence level
uncertainties, neglecting experimental errors and the cosmic variance
associated with MAX.
Points are given for both scalar and tensor contributions (solid) and scalar
only (dashed).  Each group of points is at the same $n$, the offset is for
ease of viewing only.  Higher points represent $\Omega_B=0.10$ and low
points $\Omega_B=0.03$.
\item[Figure 5] The probability distribution of the ratio of tensor to scalar
contributions to the CMB quadrupole anisotropy, compared to the ratio of
expectation values for these quantities.  Also shown is the integrated
probability distribution, useful for computing confidence limits.
\end{itemize}

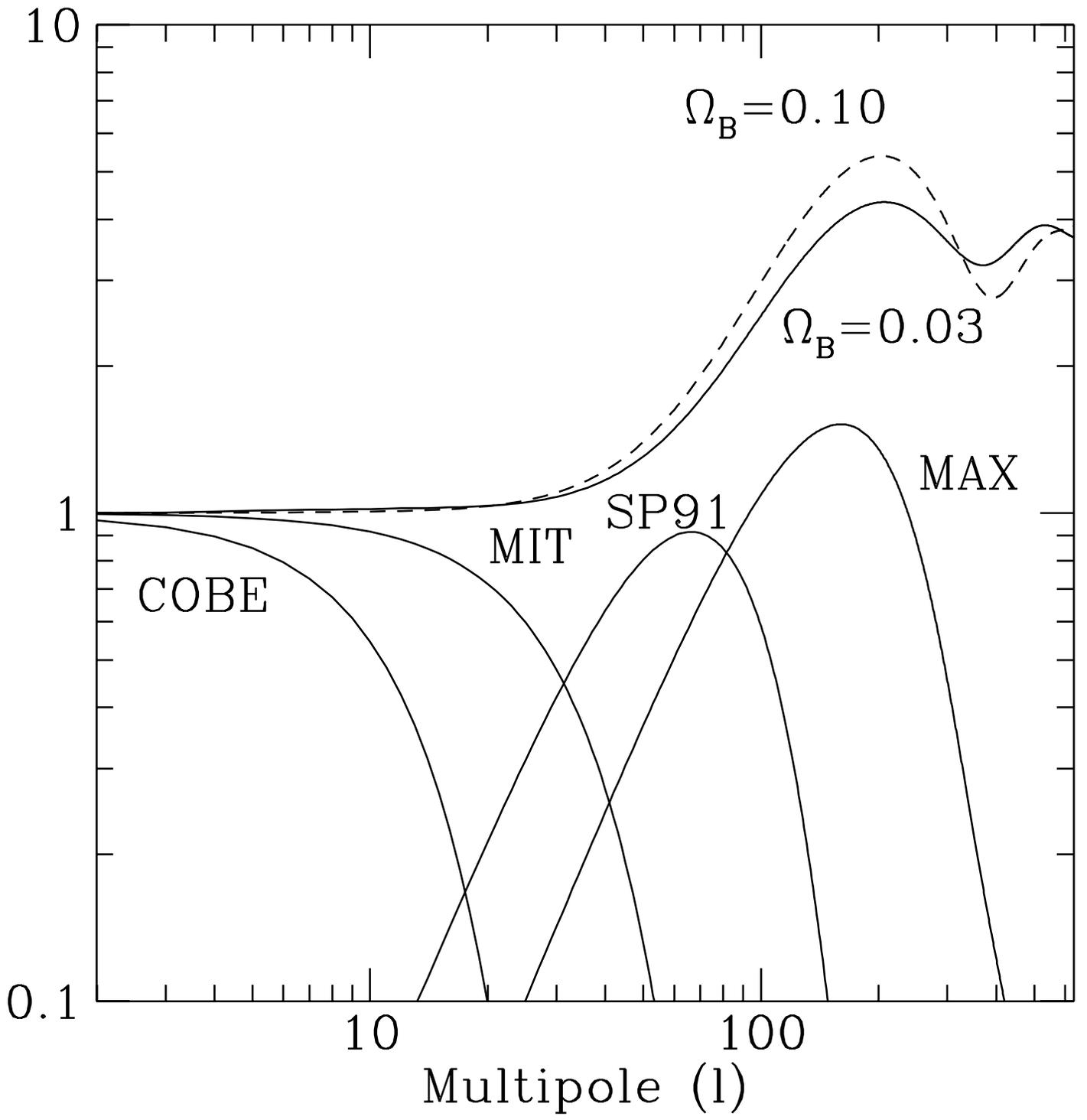
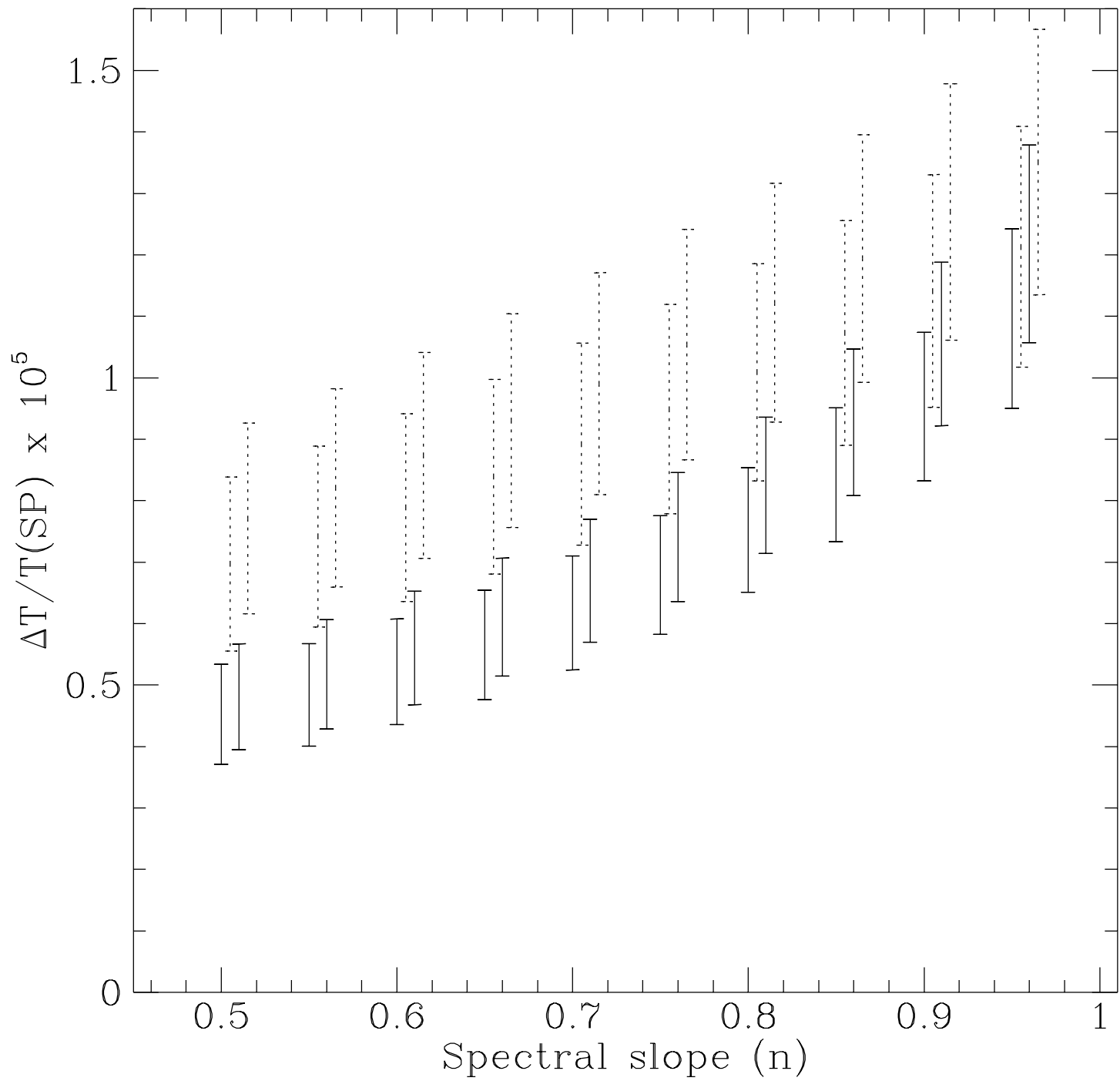
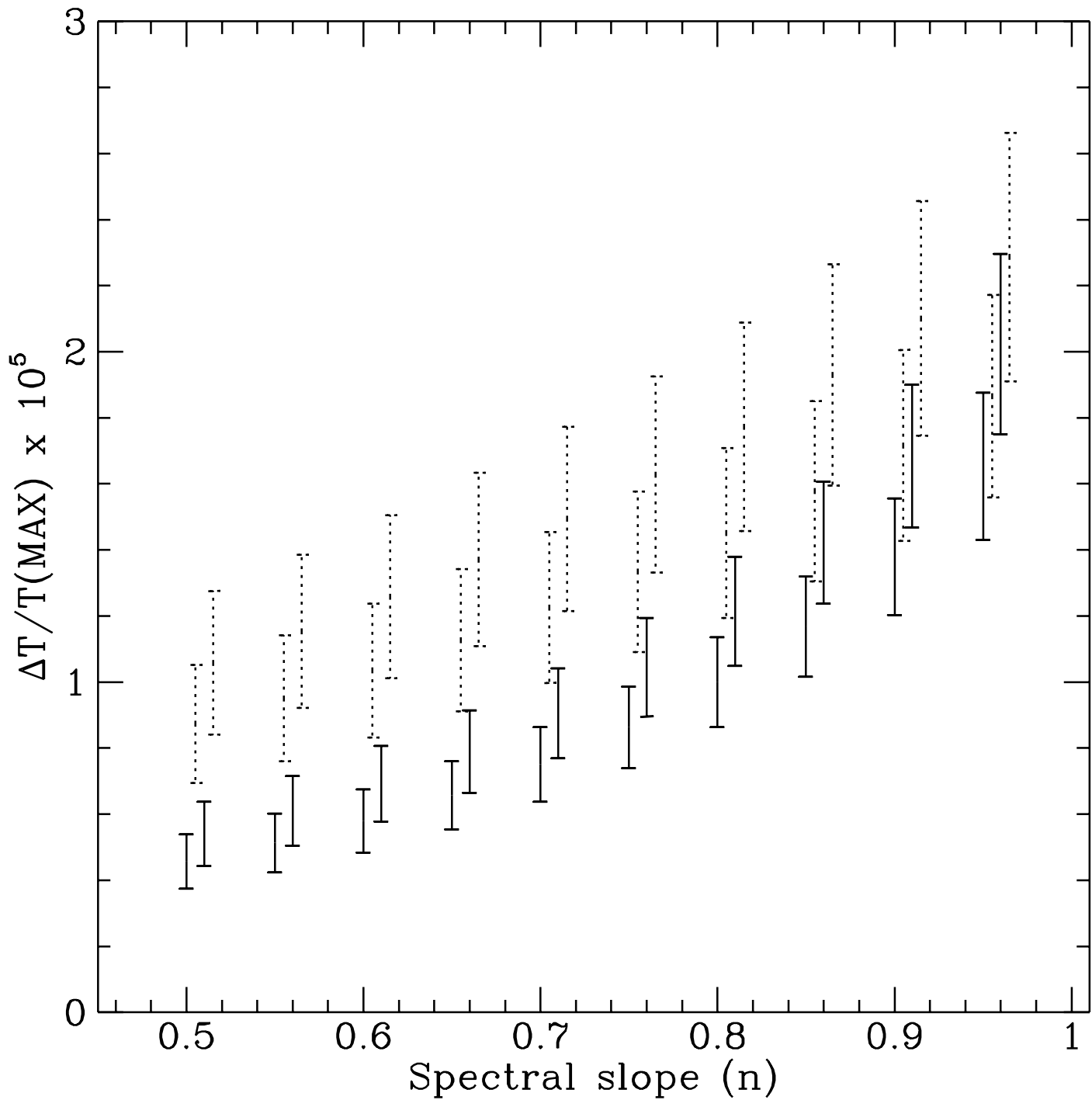
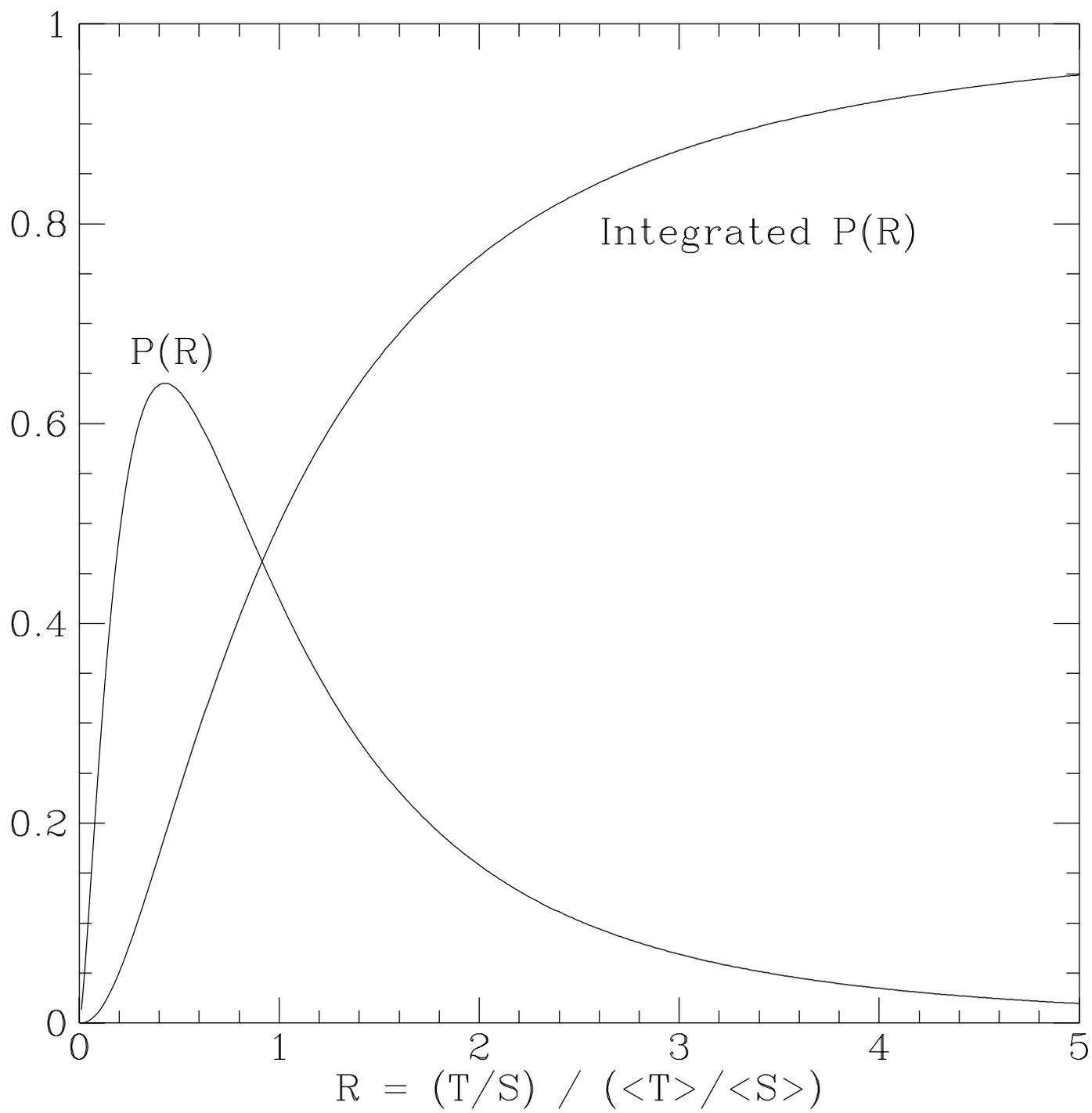
\end{document}